\documentclass[11pt,preprint2]{aastex}

\makeindex
\citeindextrue

\bibpunct{(}{)}{,}{a}{}{}

\shorttitle{VLA Survey of the CDF-S}
\shortauthors{Kellermann et al.}

\received{xxxxxxxx}
\begin{document}

\title{The VLA Survey of the Chandra Deep Field South: \\
I. Overview and the Radio Data}


\author{K. I. Kellermann, E. B. Fomalont} 
\affil{National Radio Astronomy Observatory,
520 Edgemont Road, Charlottesville, VA~22903--2475, U.S.A.}
\email{kkellerm@nrao.edu, efomalon@nrao.edu}

\author{V. Mainieri, P. Padovani, P. Rosati, P. Shaver} 
\affil{European Organization for Astronomical Research in the Southern Hemisphere (ESO),
Karl-Schwarschild-Strasse 2, D-85748} \email{vmainier@eso.org, Paolo.Padovani@eso.org,
prosati@eso.org, pshaver@eso.org}

\author{P. Tozzi} \affil{INAF Osservatorio Astronomico di Trieste, via G.B.
Tiepolo 11, I-34131, Trieste, Italy} \email{tozzi@ts.astro.it}

\author{N. Miller} \affil{Jansky Fellow, National Radio Astronomy Observatory, 520
Edgemont Road, Charlottesville, VA~22903--2475, U.S.A.and Department of Physics
and Astronomy, Johns Hopkins University, Baltimore, MD, U.S.A.}
\email{nmiller@skysrv.pha.jhu.edu}
\email{}
\email{keywords:~{radio continuum: galaxies --- X-rays: galaxies --- galaxies: active --- galaxies:starburst}}
\email{}
\email{}
\email{}

\bigskip

\acknowledgements
\bigskip
{\bf ABSTRACT:~~}We report
20 and 6 cm VLA deep observations of the CDF-S including the Extended CDF-S. We discuss the radio
properties of 266 cataloged radio sources, of which 198 are above a 20 cm completeness level
reaching down to $43~\mu$Jy at the center of the field. Survey observations made at 6 cm
over a more limited region covers the original CDF-S to a comparable level of sensitivity as the 
20 cm observations.

Of 266 cataloged radio sources, 52 have X-ray counterparts in the CDF-S and a further 37 in the E-CDF-S area not covered by the 1 Megasecond exposure. 
Using a wide range of material, we have found optical or infrared counterparts for 254 radio sources, of which 186 have either spectroscopic or photometric redshifts (Paper II).  
Three radio sources have no apparent counterpart at any other wavelength.
Measurements of the 20 cm radio flux density at the position of each CDF-S X-ray source detected a 
further 30 radio sources above a conservative 3-sigma detection limit.

X-ray and sub-mm observations have been traditionally used as a measure of AGN and
star formation activity, respectively.  These new observations probe the faint
end of both the star formation and radio galaxy/AGN population, as well as the
connection between the formation and evolution of stars and SMBHs. Both of the
corresponding gravitational and nuclear fusion driven energy sources can lead to radio synchrotron emission.  AGN and radio galaxies dominate at
high flux densities. Although emission from star formation
becomes more prominent at the microjansky levels reached by deep radio
surveys, even for the weakest sources, we still find an apparent significant contribution from low luminosity AGN as well as from star formation. 





\section{INTRODUCTION \label{intro}}

This is the first of a series of papers based on Very Large Array (VLA) 1.4 GHz (21 cm) 
and 5 GHz (6 cm) observations of the Chandra Deep Field South (CDF-S) which also 
includes the Hubble Ultra Deep Field (UDF) and the Extended Chandra 
Deep Field South (E-CDF-S).  Paper II \citep{M08a} presents the optical 
and near IR counterparts to the observed radio sources and Paper III (P. Tozzi et al, in preparation)  
their X-ray spectral properties.  Paper IV (P. Padovani et al, in preparation) discusses the source populations.  
Other papers will deal with a 1.4 GHz survey for low surface brightness sources in the CDF-S, a deeper 5 GHz survey, 
and a 1.4 GHz survey of the E-CDF-S \cite{M08}.
 
Among the most fundamental issues in astrophysics is when and how stars, galaxies,
and black holes formed and how they evolved with cosmic time.  Star and galaxy
formation is a complex process in which mergers, shocks and accretion all appear to
play important roles. Active galactic nuclei (AGN) which occur as a result of accretion
onto a massive black hole, is also an important ingredient in galaxy evolution.
Since these processes have complicated signatures throughout the electromagnetic
spectrum, multiwavelength observations are needed to determine their evolution,
dynamics and properties as a function of cosmic time.

Deep X-ray studies  \citep[e.g.,][]{2005ARA&A..43..827B,2002AJ....124.2351B, 2004AJ....128.2048B,
2003A&A...399...39R} as well as the deep radio surveys indicate an increasing
contribution to the X-ray emission at fainter flux densities from star forming activity.
Various lines of evidence suggests the simultaneous presence of AGN and star formation and a
causal relation between super massive black holes (SMBHs) and star formation. Previous deep radio surveys, 
\citep[e.g.,][]{mux05,2000A&A...361L..41G} also appear to detect radio emission caused by both AGN and star
formation in the same galaxy.  From investigations of the relationship between black hole
mass and bulge luminosity \citep{1995ARA&A..33..581K, 1998AJ....115.2285M}, or between black
hole mass and bulge velocity dispersion \citep{2000ApJ...539L...9F}, it appears that AGN
processes  occur in star-forming galaxies with both processes being driven by, or connected by,
an unknown phenomenon \citep{2004ApJ...600..580G, 1999MNRAS.308L..39F}.
Understanding the interaction of stars, galaxies and AGN phenomenon and their evolution
with time are fundamental to an understanding of galaxy formation in the early universe.  Such
studies may also shed new light on the formation and variety of the nearby, quiescent
galaxies.

Studies of cosmic evolution for radio galaxies \citep{1990MNRAS.247...19D},
optically selected quasars \citep{2001AJ....121...54F, 2000MNRAS.317.1014B}, radio
loud quasars \citep{1996Natur.384..439S, 2005A&A...434..133W}, X-ray sources
\citep{2005ARA&A..43..827B}, and star formation \citep{1998ApJ...498..106M} all
indicate a similar pattern of an increase in luminosity and/or density up to epochs
corresponding to redshifts of 1 to 2 followed by a rapid decline to current
epochs. The near simultaneity (on cosmic time scales) of these apparently very
different phenomena is curious.  What caused the cutoff in the formation of stars,
galaxies, and quasars all around the same time?  Does the presence of an AGN stimulate star formation, or do high rates of star formation lead to AGN?  Or, is there some underlying phenomenon which is responsible for both?

\subsection{Radio Studies}

The extragalactic radio source population ranges from normal galaxies with
luminosities near $10^{19} \rm ~Watts~Hz^{-1}$ to galaxies whose radio emission is as much as
$10^7$ times greater owing to regions of massive star formation and/or to an AGN.  The
population of radio sources in the sky with flux densities greater than 1 mJy is
dominated by AGN driven emission in which virtually all of the energy is
generated from the gravitational potential associated with a SMBH in the nucleus.  For these sources, the observed radio emission
includes the classical extended jet and double lobe radio source which may
extend up to Megaparsecs from the parent galaxy, as well as compact radio components
that are more directly associated with the energy generation and collimation near
the central engine of the AGN.

We will use the term ``radio galaxy'' to refer to those sources where the radio
emission is primarily from extended lobes or jets with inverted power-law spectra,
and which have optical counterparts that are of galactic dimensions and are typically associated with bright Elliptical galaxies.  We use
``AGN'' for those radio sources, including quasars, which are less than an arcsecond in size, have
flat or inverted radio spectra characteristic of an opaque synchrotron source and
whose optical counterparts appear stellar. For both the radio galaxies and AGN the power source is thought to be due to the heating of material from
infall to a SMBH ranging from $10^6$ solar masses for low luminosity
AGN to $10^8$ or more solar masses in the case of radio galaxies and quasars.  In all cases the radio emission is
believed to be due to synchrotron emission from relativistic electrons in magnetic
fields of the order of $10^{-5}$ Gauss.  

Below 1 mJy there is an increasing contribution to the radio source population from synchrotron
emission resulting from relativistic plasma ejected from supernovae of young stars
associated with massive star formation in galaxies or groups of galaxies where mergers or
interactions appear to be important \citep{win95,ric98,fom02,2005A&A...441..879C,fom06}.  However, the mix of star-formation and AGN related radio emission and the dependence on  epoch is not well determined.
Massive star-formation phenomena is a strong function of cosmic time and produces a
variety of radiation at all wavebands: synchrotron emission at radio wavelengths; warm
dust emission at infra-red and sub-mm wavelengths; shock and hot gas emission at optical
and ultra-violet wavelengths; and thermal X-ray emission from the hot gas where massive
stars are formed and from X-ray binary stars.  At least for low redshifts, the radio
emission is tightly correlated with the FIR emission \citep{1992ARA&A..30..575C, con91}. Because of the lack of absorption at
radio wavelengths, the measured radio flux density can give a good estimate of star formation
activity over $ \sim 10^8$ years at fainter levels than are probed with FIR observations. However, the radio-FIR relation has not yet been
tested out to large redshifts.  Also, \citet{2003A&A...399...39R} report a radio-
X--ray correlation for star forming galaxies.

Radio observations are particularly valuable in distinguishing between AGN and 
star-formation, as they see through the gas and dust which often surrounds both
star forming regions and AGN. At radio wavelengths, routine observations even for the faintest sources can have the sub-arcsecond resolution needed to
distinguish between radio galaxies, AGN and starburst morphology. In the later case, the radio source has dimensions about that of a galactic disk or of single regions of active star formation typically in the range 0.1 to 1 arcsecond at z $\sim$ 1.  The AGN, by contrast have dimensions of the order of $10^{-4}$ to $10^{-2}$ arcsec, characteristic of an opaque synchrotron source of 1 to 100 parsecs, while the more powerful radio galaxies show the characteristic jet-multiple lobe morphology, often along with a milliarcsec AGN component.  High angular resolution of
radio observations are also critical to unambiguously identify the host galaxy needed to establish the SED from observations over a wide range of wavelengths, to
determine redshifts and luminosities, as well as
the morphology of the host galaxy and to reveal associations in groups or
clusters.

Previous deep radio observations have been made with the VLA in e.g., the Hubble Deep Field
North \citep{ric98,ric00}, in SSA13 \citep{fom02,fom06} and other fields
\citep{1993ApJ...405..498W, mit85, cil00, 2003A&A...403..857B, 2004MNRAS.352..131S} and with the WSRT
\citep{oor85, 2000A&A...361L..41G, 2002AJ....123.1784D}.  In the southern
hemisphere, the Australia Telescope Compact Array, (ATCA) has been used to study the
Phoenix field \citep{1998MNRAS.296..839H}, the Hubble Deep Field South
\citep{2003AJ....125..465H}, and the Chandra Deep Field South and ELAIS SWIRE fields\citep{nor2006, A06}.
Combined VLA and MERLIN observations of the HDFN \citep{mux05} with a resolution of
$0.2''$ show the complexity of the microjansky radio emission suggesting a mixture of AGN and star formation which contributes to the sub-millijansky radio population. In Paper IV, we show that the well known increase in the number of sub-millijansky radio sources is apparently not entirely due to a population of star forming galaxies, and that the fraction of AGN contribution to the sub-millijansky population is greater than previously thought. 

Most of the microjansky radio sources found in these deep radio surveys are
identified with galaxies brighter than about R=26.  About 15\% are fainter; 
many are red in color and may be detected in the z and K band.  These extremely-red
objects (EROs) clearly show the presence of dust which does not affect the
centimeter radio emission, but produces the large IR and sub-mm emission and
obscures the optical emission.  The $\sim 0.2''$ position accuracy available with
the VLA is crucial in locating their X-ray and optical counterparts and in
determining the nature of the emission process.  This is particularly important
for sub-mm sources with radio counterparts, as the sub-mm observations alone have
poor positional accuracies and often ambiguous identifications.  Comparison of the
radio and the X-ray position may help distinguish whether the radio emission is dominated
by extra-nuclear star formation or by an AGN.

\subsection{X--ray Studies}

X-ray luminosities greater than about $10^{35}$ Watts are usually, but not
always, associated with AGN, QSOs, Seyferts, and other active galaxies containing
broad emission lines \citep{2003A&A...412..689P}. However, not all AGN are
observed as strong X--ray sources in the Chandra or XMM bands.  Some may be heavily obscured \citep{DRRP05} or Compton thick, a phenomenon
which occurs for absorbing column densities of equivalent hydrogen atoms assuming solar metalicity are larger than $\sim 10^{24}$~cm$^{-2}$, when the reflected
component is expected to dominate the X-ray spectrum.  Intrinsic absorption of
less than about $10^{22}$ to $10^{24}$~cm$^{-2}$, corresponds to Compton thin sources where
the X-ray emission is usually easily observed.  At least 80 percent of these
sources in the CDF-S are associated with narrow emission line type II objects.  Starting
nominally at $1.5 \times 10^{24}$~cm$^{-2}$,  the Comptonization effects start to be
relevant, direct emission is strongly suppressed, and therefore we expect that most of the X--ray emission will be due to
reflection from a slab of cold material.  In this case the observed X-ray emission
may be only of the order of a few percent of the intrinsic power, and such objects are
therefore expected to be only weak X-ray sources.  According to \citet{2005MNRAS.357.1281W}
half of the hard X-ray background (XRB) above 5 kev remains unresolved and is shown to be consistent with the XRB being due to
Compton-thick sources at $z \sim 1$.  This means that XMM
and Chandra observations are missing a large part of this population of sources. Radio observations
may be sensitive to these Compton-thick sources with suppressed X-ray emission; for example,
\cite{2005Natur.436..666M} use radio and FIR data to identify highly absorbed
X-ray weak AGN.

X-ray sources with luminosities in the range $\sim 10^{34}$ Watts to $10^{35}$
Watts include a mixture of low luminosity AGN and star forming galaxies which
cannot be distinguished from X-ray observations alone.  Although the weaker X-ray sources
appear to be increasingly dominated by star
formation, unlike the weak radio surveys, the deep X-ray surveys do not show any
evidence for mergers or other interactions \citep{2005ARA&A..43..827B}.  

\subsection{The Chandra Deep Field South}

The CDF-S
is probably the most intensely studied region of sky with extensive multiwavelegth
observations using the most powerful space and ground facilities, and is uniquely
suited for studies of the co-evolution of star formation and AGN. High sensitivity
X-ray observations are available from Chandra
\citep{2002ApJS..139..369G,2005ApJS..161...21L}.  The GOODS \citep{2004ApJ...600L..93G} and GEMS \citep{RBB04} multiband
imaging programs using the HST Advanced Camera for Surveys (ACS) give sensitive
high resolution optical images over the CDF-S field \citep{B06}. Ground based imaging and spectroscopy are
available from the ESO 2.2m and 8m telescopes, and IR observations from the Spitzer Space Telescope. Also, located near the center of the CDF-S, is the Hubble Ultra Deep Field (UDF) with
its unprecedented sensitivity in four optical bands reaching a limiting magnitude
as faint as 29 at 775 nm.

In this paper we report new radio observations of the CDF-S, including the Hubble UDF, and the E-CDF-S,
made with the NRAO Very Large Array (VLA)
at 1.4 GHz (20 cm) and 5 GHz (6 cm).  The effective angular resolution was 
$3.5''$ and  minimum rms noise as low as $8.5~\mu$Jy per beam at 20 cm and $7~\mu$Jy per beam at 6 cm.  These deep
radio observations complement the larger area, but less sensitive lower resolution
observations of the CDF-S discussed by \cite{nor2006} and \cite{A06}.

\section{The VLA OBSERVATIONS\label{VLA}}

We observed the CDF-S for 50 hours at 20 cm primarily in the BnA configuration
in October 1999 and February 2001, and for 32 hours at 6 cm primarily in the C
and CnB configurations between June and October 2001. At both wavelengths, we referred
the observations to the phase and amplitude calibrator J0340-213 located at
$\alpha=03^h40^m35.6079^s$ and $\delta=-21^\circ 19' 31.172''$; (J2000). 
This absolute position with respect to the International Celestial
Reference Frame has been measured with the Very Long Baseline Array to an accuracy
of better than 0.01'' \citep{ma98}.  The relative positional accuracy obtained
between J0340-213 and the CDF-S radio field depends on their angular separation,
the a-priori calibration of VLA astrometric parameters, and weather conditions.
Based on many astrometric VLA observations, the relative positional accuracy at
1.4 GHz in the B-configuration is believed to be better than 0.1'' and this is the minimum
position error we adopt in each coordinate for any source in the CDF-S radio
catalog.

\subsection{The 20 cm observations and data reduction}

The field center of the VLA 20 cm image is at R.A.=$03^h32^m28.0^s$, DEC=
$-27^\circ48'30.0''$ (J2000).  The observations were taken at two frequencies,
1.365 GHz and 1.435 GHz, each with dual circular polarization.  We observed in 15
minute segments, alternating between 12 minutes on the CDF-S and 3 minutes on the
calibrator J0340-213, and a linear interpolation of the gain and phase calibration,
determined from the calibrator was applied to the CDF-S
observations.

The flux density scale was determined by daily observations of 3C 286
assuming a flux density of 16.38 Jy and 15.75 Jy derived from
\cite{1977A&A....61...99B} at 1.365 GHz and 1.435 GHz respectively. In order to
minimize the image distortion and loss of sensitivity over the full area of sky covered
by the Chandra image, the data in each frequency/polarization channel were split
into 7 channels, each of 3.125-MHz bandwidth, and the data were sampled every 5
seconds.  This resulted in a loss of sensitivity at the field center by a factor
of 0.71, compared with using the full VLA nominal continuum bandwidth of 50 MHz, but gives better average sensitivity over the full CDF-S field.  A few percent of the data from each
observing session was deleted for occasional short periods of interference, during
periods of known technical problems, for telescope shadowing, and during inclement
weather conditions.

Since the 20 cm sensitivity of the VLA is significantly degraded at low
elevations, and due to the low declination of the CDF-S field, we observed each day
for only 5 or 6 hours roughly centered on the meridian.  Antenna based gain and
phase fluctuations were determined every 15 minutes from the calibrator
observations.  The gain variations between successive calibrator observations was
generally less than 2\%, apart from the decreasing loss of sensitivity at
elevations below $15^\circ$ of about 30\% caused by the additional noise from
ground emission.  The phase fluctuations between calibrator scans were typically
$2^\circ$ to $10^\circ$ on the short to the long spacings, respectively.  During
periods of inclement weather and at sunrise or sunset when ionospheric activity
can be large, phase changes of $15^\circ$ over a few minutes occurred on baselines
longer than about 10 km. 

   The self-calibration and imaging of the VLA data were complicated by two
factors: First, sources more than about $5'$ from the field are distorted by the
finite bandwidth of each 3 MHz channel (chromatic aberration), by the finite data
sampling interval, and by sky curvature. Secondly, the non-circularity of the VLA
primary beam produces an apparent variability of sources because of the relative
rotation between the plane of the sky and the primary antenna sensitivity pattern
during an observation day.  Both of these effects produce artifacts near bright
sources of about one percent.  Since the brightest peak flux density in our
1.4 GHz image is 37 mJy, these artifacts can be well above the rms noise level of
$8~\mu$Jy.  Thus, special imaging and self-calibration techniques were necessary
to reduce these artifacts to the level of that expected from thermal noise alone.

After applying the calibrations, based on the measurements of J0340-213, a CLEAN
(deconvolved) image of radius $40'$ centered on the CDF-S was made using the AIPS
task IMAGR.  We used a pixel size of $0.5''$ and a resolution of 3.5 arcsec to
match that of the 6 cm image described in Section \ref{6cm}.  The CDF-S field was
covered with 30 overlapping images each $512\times 512$ pixels in size.  The individual image size
was chosen to limit the distortion, caused by the sky curvature, to under $5\%$
for a source at the edge of each image.  We also made smaller images around the 14
brightest sources outside the main field but within $90'$ of the field center 
that were stronger than 1 mJy in the uncorrected image.
Because of the relatively accurate calibration of the data using J0340-213 the
initial radio images were of good quality, although the expected artifacts at the
$50~\mu$Jy level were clearly associated with the strongest sources.

     For the next calibration step, the standard self-calibration technique (CALIB
and IMAGR in AIPS) was used to improve the calibration of the CDF-S data.  Because
of the large number of baselines (351) compared with the number of telescopes
(27), it is possible to improve the telescope calibration using a source model,
which is in fact the first-pass clean image \citep{cor89}.  With the self-calibration algorithm, each telescope residual phase was determined every four
minutes for each polarization and frequency. The signal-to-noise in each of these
intervals was about 30:1, so robust solutions were obtained.  As expected, these
residual phase determinations were generally less than $5^\circ$, although there
were times when they were as large as $30^\circ$; these coincided with periods
when there was significant phase change in the original calibration with
J0340-213\footnote{The residual amplitude calibration was also determined;
however, it was close to unity and significant departures were indicative of
periods of poor data quality, rather than improved gain calibration.}. Because
these residual phases were primarily associated with the troposphere and
ionosphere above each telescope, the four phase determinations---derived from two
frequencies and two polarization---agreed to within a few degrees. During the rare
periods when there was not good agreement among the four determinations, or lack
of reasonable continuity among the telescope solutions, or further inspection
indicated other problems, these data were deleted.  With the improved phase
calibration and additional data editing, new cleaned images were then made.

The dynamic range of deep VLA observations are limited by artifacts
from the strongest sources caused by their apparent variability as the
sky rotates with respect to the non-circular VLA primary beam pattern or by small but variable pointing errors.
These artifacts, generally about 0.1\% of the brightness peak, are not
removed with the clean deconvolution.  We first imaged and cleaned the
the data in 2-hour observing blocks for each of the two frequencies
separately for each day.  During this 2-hour period, the apparent source
variability was minimal.  This produced 6 clean images per observing
day.  We then subtracted the contribution from the 14 bright sources
from each 2-hour time segment and frequency directly from the data
base to form a residual data base with the apparent flux density
variations of the strong sources removed.  Finally, we imaged and
deeply cleaned this residual data in the 30 overlapping $512\times
512$ images to the $24~\mu$Jy level (2.8-$\sigma$) to produce the
final image.  The cleaned images of the 14 subtracted sources
(averaged over all separate time-sliced images) were added to the
final image so that all of the emission from all sources is included
in the final image

The final image (uncorrected for the primary beam attenuation) has an rms noise of
$8.5~\mu$Jy, constant over the entire image area.  No significant
increase in rms, caused by possible source confusion or residual faint
sidelobes from the higher density of sources, was observed near the
field center.  The formal completeness level of the image was taken to
be at a peak image flux density of $43~\mu$Jy per beam, or 5 times the
rms noise level.  Based on the almost Gaussian noise characteristics
of the image, there is less than a 1\% chance that an apparent
source in the image at or above this level is merely due to a
noise or instrumental fluctuation.  Hence, we expect no more than two of
the cataloged sources above this detection limit to be false detections.  The
largest negative peak seen is $-43~\mu$Jy which supports our chosen
completeness level.  This does not mean that all sources with a true
peak flux density of $43~\mu$Jy near the center of the field are
included in the catalog.  The effect of the $8.5~\mu$Jy rms noise is to cause an uncertainty in quoted flux density which will cause some weaker sources to appear above this limit, and remove some
stronger sources from the catalog. This causes a non random bias which affects the source count, but as explained in Section 7, the count can be corrected for these biases.

\begin{figure}[ht] 
\begin{center}
\special{epsfile=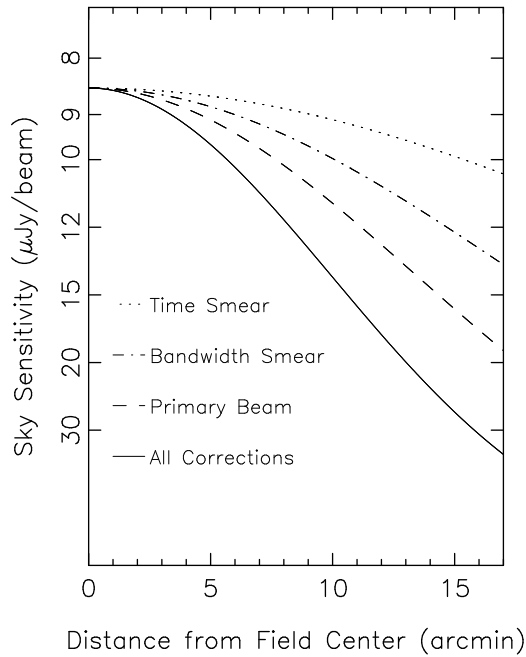,width=4in} 
\resizebox{0.9\hsize}{!}
{\includegraphics[trim=0cm 0cm 0cm 0cm,clip]{f1.eps}} 
\end{center}
\caption{\label{sensitivity} 1.4 GHz (20 cm) rms noise in microjanskys as a function of distance
from the field center. } 
\end{figure}

To obtain the true sky image from the VLA image, three effects must be removed.  First, due to the primary beam shape of the 25m VLA antenna elements, the sky sensitivity of the
observations decreases with radial distance from the field center, with a full-width at half-power (FWHP) of $31.5'$. This primary beam radial attenuation
function has been measured to an accuracy of 2\% and is nearly circular. For this correction, we used the
experimentally-measured sensitivity, and the major effects from the beam non-circularity were removed as described above.   Second,
because of the chromatic aberration, there is an addition smearing in the image
which broadens each source in the radial direction.  Correction for this attenuation
can be applied to obtain the flux density and the source angular size or limit.
Finally, there is a slight decrease in resolution and coherence caused by the
finite sampling time and the effective sensitivity loss.  The combined attenuation
is then used to calculate the rms noise as a function of distance from the field
center and is shown in Fig.~\ref{sensitivity}.  At the most distant parts of the
CDF-S from the field center, the sky sensitivity is 54\% of that at the field
center so that the rms noise level is $16~\mu$Jy, and the catalog is thought to be  ``complete'' over the entire 
CDF-S for peak flux densities greater than $80~\mu$Jy.

\subsection{The 20 cm source list}

   An initial list of radio sources with SNR greater than 4.5 times the rms noise
was generated from the region within $17'$ of the field center. The parameters of
most sources in the radio image were determined by fitting the brightness
distribution to a single elliptical Gaussian component.  The six independent
parameters for each component were the peak flux density, the x- and y-
coordinates, and the major, minor axes and orientation of the elliptical
component.  The integrated flux density is equal to the peak flux density times
the component area divided by the beam area (as modified by instrumental effects).
If the source was not appreciably resolved (the difference between the peak flux
density and the integrated flux density was less than two times the rms noise in
the image), then the average of the measured peak and integrated flux densities
were assigned as the final integrated flux density, with the appropriate diameter
limits.

For sources with an image peak flux density less than $85~\mu$Jy (10 times the rms
noise), non-linear bias of the fitting and deconvolution process, due to receiver
noise, confusion bias and residual side-lobe contamination, becomes important
\citep{1984A&AS...58....1W}.  Using simulations as a guideline of these biases on
the determination of radio angular diameters, we adopted a conservative approach,
and do not quote sizes unless the probability that a radio source is resolved (not
consistent with a point source) was greater than 85\%. Otherwise, we only placed
an upper limit to the angular size.

The density of radio sources is sufficiently high that in some cases it was
difficult to determine if a blend or near blend of two radio sources is in fact
one extended radio source or two close, but independent, radio sources.  This is
particularly true when considering weak sources somewhat below our completeness
level, but located close to a somewhat stronger source.  Thus, before making the
final radio catalog, we compared the radio emission and optical emission and used
the identification information as an aid in defining the individual radio source
specifications.

\begin{figure}[ht] 
\begin{center}
\includegraphics{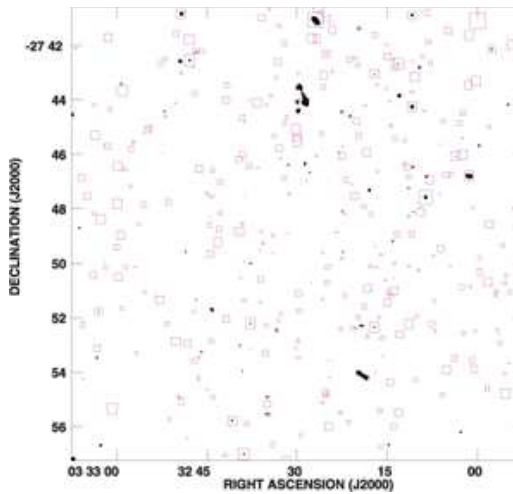,width=4in} 
\resizebox{0.9\hsize}{!}
{\includegraphics[trim=0cm 0cm 0cm 0cm,clip]{f2.eps}} 
\end{center}
\caption{\label{VLA_radio}The location of the VLA 20 cm radio sources are shown in black with the location
of each Chandra X-ray source shown by a purple box.  The size of each box is
proportional to the log of the total X-ray flux.} 
\end{figure}

The final cleaned sky 20 cm radio image of the CDF-S, covering a 16' by 16' region with
$3.5''$ resolution, is shown in Fig.~\ref{VLA_radio}.  The location of all
CDF-S X-ray sources is shown as overlaid squares. The rms noise is $8.5~\mu$Jy at the
field center and increases smoothly away from the center according to Fig.
\ref{sensitivity} except near a few of the very brightest sources where
there is some residual side lobe response. 

\subsection{The 6 cm data\label{6cm}}

\begin{figure}[ht] 
\begin{center}
\special{epsfile=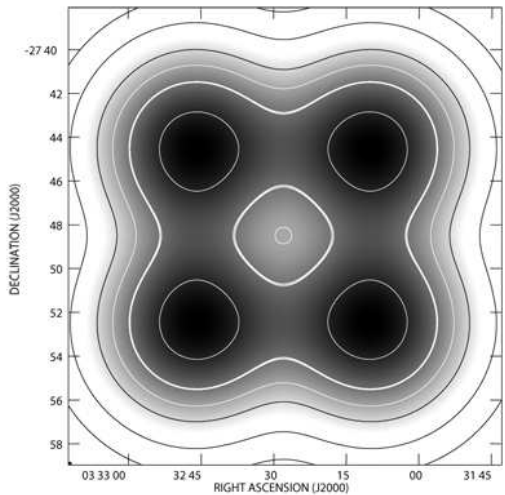,width=4in} 
\resizebox{0.9\hsize}{!}
{\includegraphics[trim=0cm 0cm 0cm 0cm,clip]{f3.eps}} 
\end{center}
\caption{\label{6cmsens}  Contour map showing the rms noise on the 5 GHz VLA image as
a function of position in the sky.  Contour levels are at 8, 10, 12, 15, 25, and
50 $\mu$Jy per beam rms.  The contour at $10~\mu$Jy per beam is shown bold.  The
grey scale goes from $7.3~\mu$Jy (black) to $25~\mu$Jy per beam rms.} 
\end{figure}

The 6 cm observations were made in two 50 MHz bands centered at 4.835 and 4.885 GHz each with both LHC and RHC polarization primarily in CnB VLA configuration which gives a matched resolution to the BnA configuration used for the 20 cm observations.  At 6 cm, the primary beam of the VLA is only 7.5 arcmin FWHP, so in order to cover
the entire CDF-S we used four pointing centers located at the corners of a square
spaced 8 arcmin apart symmetrically spaced around the same pointing center used for the 20 cm observations.  This gave a reasonably uniform response over most of the CDF-S.
Because the source density at 6 cm is less than at 20 cm and due to the smaller
primary beam, the straight forward normal VLA/AIPS calibration procedure was sufficient to give
noise limited images.  The rms noise is less than $15~\mu$Jy per beam over most of the
CDF-S, but due to the mosaicing observing method used, the noise varies across the
field from $7~\mu$Jy per beam near each of the four fiducial points to $11~\mu$Jy per beam at the
center of the CDF-S to near $20~\mu$Jy per beam close to the extreme edges of the CDF-S. In
Fig. \ref{6cmsens} we show a contour map of the rms noise in the 6 cm VLA image of the CDF-S.  The 6 cm flux density was determined for all sources with a peak map flux density greater
than 5 times the local rms noise, and in addition, for each cataloged 20 cm source whose 6 cm flux density did not exceed the 5-sigma noise cutoff, the 5 GHz flux density was measured at the position
of the 20 cm source.  There were no
sources detected above 5-sigma cutoff at 6 cm which are not included in the 20 cm catalog.

\section{THE RADIO CATALOG}

A total of 266 radio sources were cataloged at 20 cm.  Of these, 198 are believed to form a
``complete sample'' with a peak SNR greater than 5 and located within 15 arcmin of the field
center.  The corresponding flux density limit ranges from 43 $\mu$Jy at the field center to
125 $\mu$Jy near the field edge.  Table~\ref{catalog} presents the radio
source list and is organized as follows:

\footnotesize \vskip 0.3cm

Col. 1~~The Source Name following the IAU definition.  An entry "N" indicates that the source is not in the complete 20 cm catalog as defined in the text.

Col. 2~~A running number we have found convenient to use. Sources given in bold
have multiple components. This entry is followed by an entry for each of the
individual components denoted by capital letters.  For these sources the position
given is that of the "core" component associated with an optical counterpart or in
a few cases where there is no observed "core," which we believe should be
identified with the "core."

Col. 3~~The Right Ascension and Declination (J2000) and their rms errors determined from the 1.4 GHz data.

Col. 4~~The signal to noise ratio on the 20 cm image

Col. 5~~The 20 cm integrated sky flux density in and rms error in microjanskys

Col. 6~~The overall angular size in arcsec measured at 20 cm.

Col. 7~~The flux density at 6 cm and rms error in microjanskys.  In addition to those
sources selected as having a peak flux density greater than 5 time the rms noise,
we have also measured the 6 cm flux density at the position of each of the
cataloged 20 cm sources.

Col. 8~~The spectral index between 6 and 20 cm and the rms error defined by S $\propto
 (freq)^{-\alpha}$.

Col. 9~~The redshift of the most likely optical counterpart.  Where available, we give spectroscopic redshifts; otherwise we show photometric values.  See Paper II

Col. 10~~The R magnitude of the most likely optical counterpart. In the case of multicomponent  sources, where possible, we show the magnitude with the associated radio component. See Paper II

Col. 11~~The Chandra ID taken from \cite{2002ApJS..139..369G} or from the E-CDF-S
data of \cite{2005ApJS..161...21L}.  Sources from the E-CDF-S have the prefix E. In the case of multicomponent  sources, where possible, we show the XID with the associated radio component.

Col. 12~~The log of the Chandra full band flux in the range 0.5 to 10 keV in $\rm ergs~cm^{-2} ~s^{-1}$. In the case of multicomponent sources, where possible, we show the value with the associated radio component.

Col. 13~~The log of the Chandra full band luminosity in  $\rm ergs~sec^{-1}$

Col. 14~~The log of the 1.4 GHz (20 cm) rest frame radio luminosity in $\rm W~Hz^{-1}$.  K-corrections were applied based on the spectral index, if available, as given in Col. 8; otherwise assuming a value of 0.7.

\normalsize

For sources with multiple components we give on the first line for each source the centroid position and total radio flux density and, where known, the corresponding redshift of the optical counterpart and the total 1.4 GHz radio luminosity.  This is followed by separate entries for each radio component showing the component radio position and flux density along with the corresponding optical magnitude, X-ray identification and X-ray flux.  For those multiple component sources where we were unable to uniquely associate the optical or X-ray counterpart with a specific radio component, we give the optical and X-ray data on the line containing the integrated radio data.

\section{Ancillary Data \label{other_data}} 

\begin{figure}[ht] 
\begin{center}
\special{epsfile=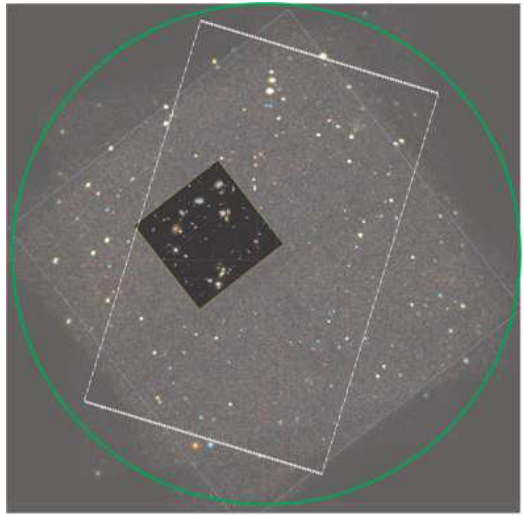,width=4in\textwidth} 
\resizebox{0.9\hsize}{!}
{\includegraphics[trim=0cm 0cm 0cm 0cm,clip]{f4.eps}} 
\end{center}
\caption{\label{overlays} Chandra image of the CDF-S showing the location of the
exposures in other wavelength bands as follows: green circle -- 30 arcmin diameter VLA 30 percent response area; solid purple squares--
Extended Chandra Deep Field South (E-CDF-S); solid grey -- ESO Imaging Survey Wide
Field Imager (EIS/WFI); white dot--dashed -- GOODS ACS field; solid yellow --
Hubble Ultra Deep Field (UDF). } 
\end{figure}

Galaxies with active star formation are often seen as luminous sub-mm \citep
{2002PhR...369..111B, 2003ApJ...585...57C} and FIR sources, and they may be
classified as ULIRGS or EROs.  Quasars and AGN are frequently seen as sources of
hard X-ray emission, and like radio emission, X-ray's penetrate the gas
surrounding obscured quasars and AGN.  However, \citet{2003A&A...399...39R} argue that both 
hard and soft X--ray emission are seen from star formation as well, although X--ray luminosities 
greater than $10^{42} ~\rm ergs~cm^{-2} ~s^{-1}$ are mostly associated with an AGN.


\subsection{\it The X-ray Data \label{xray}}

The CDF-S was the target of a 1 Megasecond Chandra observation \citep {2002ApJS..139..369G}
reaching limiting fluxes of ${\rm 5.5 ~x~ 10^{-17}~ erg~ s^{-1}}$ ${\rm cm^{-2}}$
in the [0.5-2keV] {\it soft}~band and $${\rm 4.5~x~10^{-16}~erg~ s^{-1}~ cm^{-2}}$$ in the
[2-10 keV] {\it hard}~band.  The 1 Megasecond catalog, which covers an area 16 arc minutes on a side, includes 338 X-ray sources.
Improved positions and fluxes were later published
by \cite{2003AJ....126..539A}.  The new Chandra positions, which we have used in our analysis, are corrected 
for the offset between the original X-ray and optical astrometry noted by \cite {2002ApJS..139..369G}.  A further 20 X-ray sources in the CDF-S were isolated by \cite{T06}. More recently, an additional 1 Megasecond exposure of the CDF-S was obtained bringing the total exposure to 2 Megaseconds; however that data was not available for use in this paper.


\cite{2005ApJS..161...21L} have reported the results of additional
observations at four separate contiguous pointing centers each of 250 ksec duration.  This
Extended CDF-S (E-CDF-S) field was designed to minimize the effects of cosmic
variance and covers an area 32 arc min on a side with flux limits of ${\rm
1.5~x~10^{-16}~ erg~ s^{-1}~ cm^{-2}}$ in the soft band and ${\rm
1.0~x~10^{-15}~ erg~ s^{-1}~ cm^{-2}}$ in the hard band.  As for the 1 Msec survey,
we have used a new analysis of the X-ray data to determine more accurate
fluxes in the E-CDF-S.  The E-CDF-S has
also been studied using XMM-Newton by A. Streblyanska et al. (in preparation).
with an exposure time of 500 ksec to give X-ray spectroscopic data for sources in
this field.

The X-ray fluxes and luminosities used in this paper were obtained by using conversion factors appropriate for a power law spectrum with $\Gamma = 1.4$ and the measured net count rates.  In Paper III (P. Tozzi et al., in preparation) we will present more accurate X-ray flux and luminosity estimates using spectral fits of the X--ray spectra for all the sources with a measured redshift.\footnote{The x-ray data for the CDF-S may be found at: \url{\\http://www.mpe.mpg.de/$\sim$mainieri/cdfs\_pub/} and for the E-CDF-S at (http://www.astro.psu.edu/~niel/ecdfs/ecdfs-chandra.html).}

\subsection{\it The Optical Data \label{optid}}

The VLA field has also been imaged by HST, Spitzer, the ESO 8m VLT and 2.2m telescopes as
described below.

GOODS/ACS: As part of the GOODS program, an area of 160 arcmin$^{2}$ in the center
of the CDF-S has been imaged with the Advanced Camera for Surveys on board HST
\citep{2004ApJ...600L..93G}. Observations have been made in four filters: F435W
(b), F606W (v), F775W (i), F850LP (z) down to the following limiting magnitudes:
26.7, 26.7, 25.9, 25.6 (AB, 5$\sigma$, 1 arcsec area).\footnote{See 
http://www.science.edu/science/goods/ and \cite{2004ApJ...600L..93G}}

UDF/ACS: The 200 by 200 arcsec area of the Hubble Ultra Deep Field has been imaged
with ACS in the same four filters used for GOODS but to fainter limiting
magnitudes: 28.7, 29.0, 29.0, 28.4 (AB, 5$\sigma$ for 0.2 arcsec aperture).\footnote{See
http://www.stsci.edu/hst/udf.}

GEMS/ACS: Almost the entire area covered by the VLA observations and the E-CDF-S
has a shallower ACS imaging as part of the GEMS project. Two filters have been
used: F606W (v) and F850LP (z).  The area covered is about 800 arcmin$^{2}$.\footnote{See: http://www.mpia.de/GEMS/gems.htm}

VLT: Imaging in two bands has been obtained as part of the follow-up program of
the X-ray observations: R band with VLT FORS down to 26.4 (AB, 5$\sigma$ for 2.0
arcsec aperture) in an area of 300 arcmin$^{2}$ and I band with VLT FORS down to
26.3 (AB, 5$\sigma$ for 2.0 arcsec aperture).

EIS/WFI: using the Wide Field Imager (WFI) at the 2.2 m telescope in La
Silla (Chile) as part of the ESO Imaging Survey (EIS), an area of 0.44 deg$^{2}$
(including the full VLA region) has been observed in five filters: U (26.0), B
(26.4), V (25.4), R (25.5) and I (24.7) (AB, 5$\sigma$ for 2.0 arcsec apertures).\footnote{See http://www.eso.org/science/goods/imaging/products.html}

\begin{figure}[ht] 
\begin{center}
\includegraphics{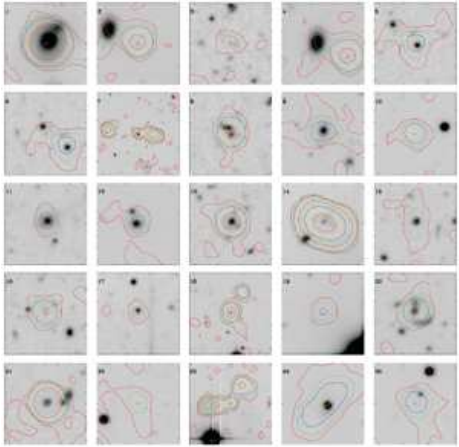,width=\textwidth} 
\resizebox{0.8\hsize}{!}
{\includegraphics{f5.eps}}
\end{center}
\caption{\label{WFI_IDs} Radio contours based on the 1.4 GHz VLA map uncorrected
for gain variation over the field and overlayed on WFI-R band images.  Each image is 20 arcsec on a side except for RID~7 which is 70 arcsec, RID~18 which is 40 arcsec, and RID~23 which is 50 arcsec.  The location of the radio centroid is shown by a red plus sign 
and any X-ray counterpart by a blue cross.  
The radio contour levels are at 20, 50, 100, 250, 500, 1,000, 2,500, 5,000, 10,000, and 20,000 $\mu$Jy per beam.}
\end{figure}

In Figures 5 through 15, we show the WFI-R band image for each radio field with the
radio contours overlaid and with the position of the X-ray counterpart, if any,
also indicated. We show only R band images and tabulate in Table~\ref{catalog} only the R band magnitudes.  However, since the R band counterpart is sometimes very faint, in assigning counterparts we have used other optical and near IR material as discussed in Paper II.  Mainieri et al. (2008b, in
preparation) discuss the multiband properties of the optical/IR counterparts to
the cataloged radio sources.

\begin{figure}[h] 
\begin{center} 
\includegraphics{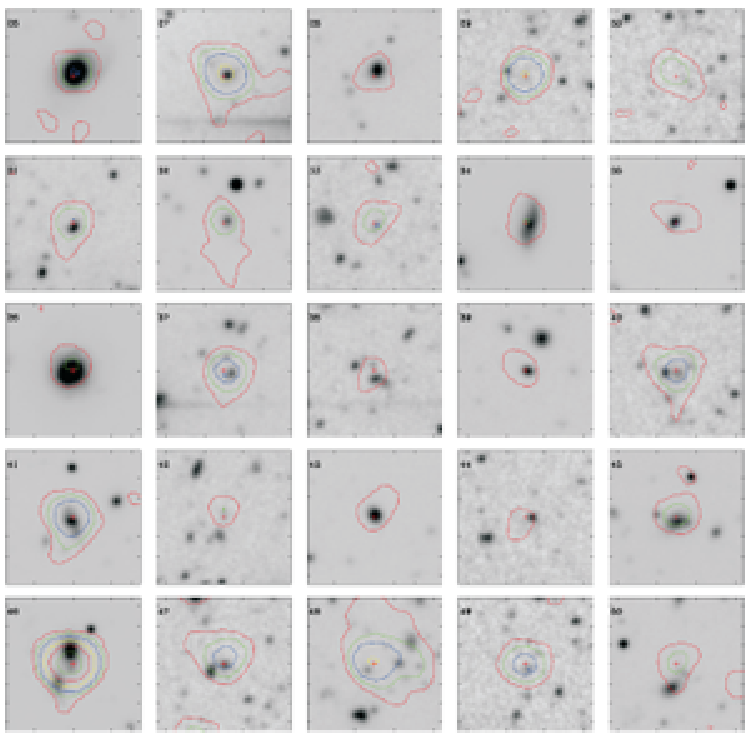,width=\textwidth}
\resizebox{0.8\hsize}{!}{\includegraphics{f6.eps}}
\end{center} 
\caption{\label{WFI_IDs} See Figure 5 caption.} 
\end{figure}

\begin{figure}[h] 
\begin{center} 
\includegraphics{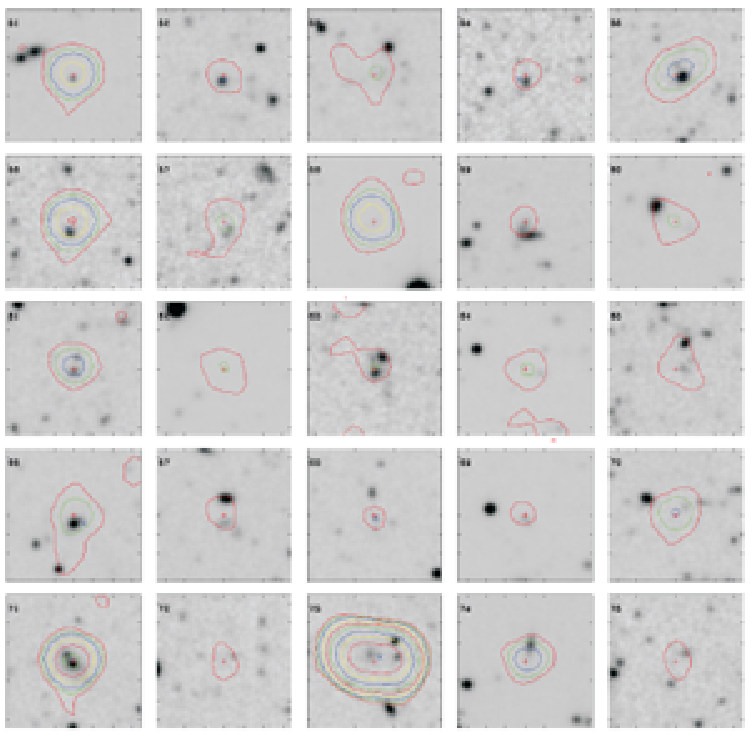,width=\textwidth}
\resizebox{0.8\hsize}{!}{\includegraphics{f7.eps}}
\end{center}
\caption{\label{WFI_IDs} See Figure 5 caption. } 
\end{figure}

\begin{figure}[p] 
\begin{center} 
\includegraphics{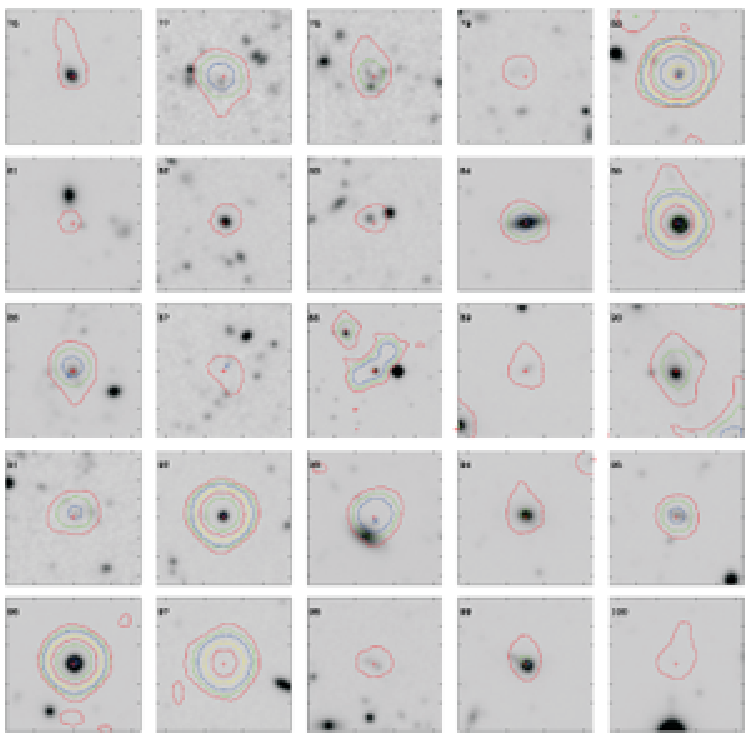,width=\textwidth}
\resizebox{0.8\hsize}{!}{\includegraphics{f8.eps}}
\end{center} 
\caption{\label{WFI_IDs} See Figure 5 caption. RID~88 is 40 arcsec on a side. } 
\end{figure}


\begin{figure}[p] 
\begin{center}
\includegraphics{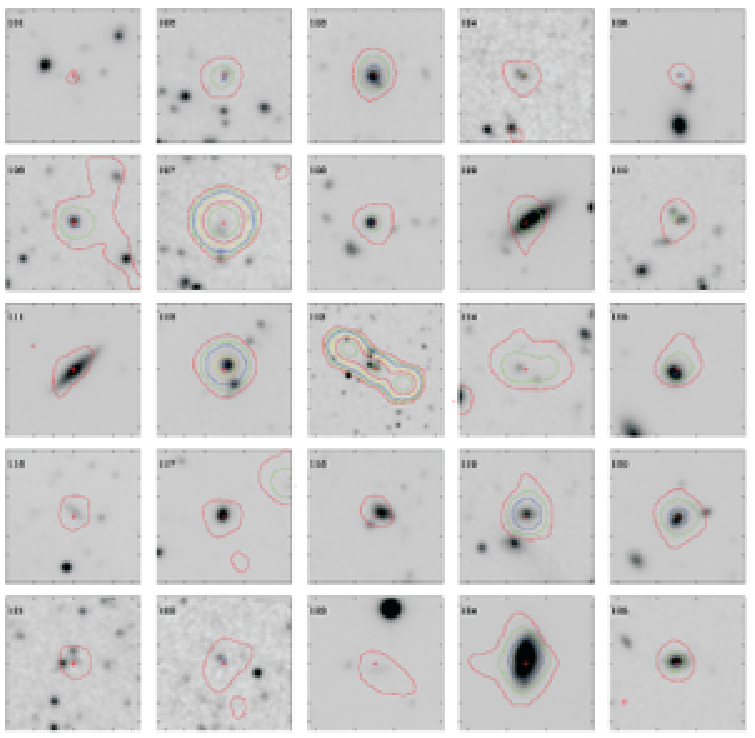,width=\textwidth} 
\resizebox{0.8\hsize}{!}
{\includegraphics{f9.eps}} 
\end{center}
\caption{\label{WFI_IDs} See Figure 5 caption. RID~113 is 40 arcsec on a side. } 
\end{figure}

\begin{figure}[p] 
\begin{center} 
\includegraphics{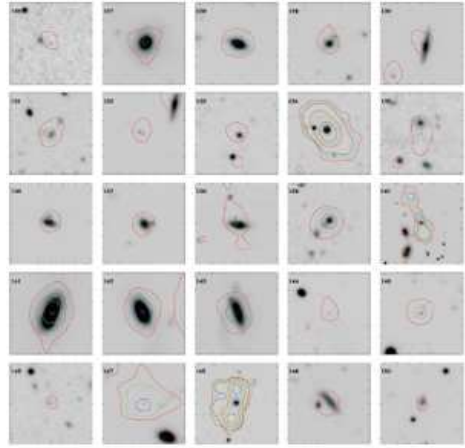,width=\textwidth}
\resizebox{0.8\hsize}{!}{\includegraphics{f10.eps}}
\end{center} 
\caption{\label{WFI_IDs} See Figure 5 caption. RID~134 is 30 arcsec on a side, RID~140 is 80 arcsec, and RID~148 is 40 arcsec. } 
\end{figure}

\begin{figure}[p] 
\begin{center} 
\includegraphics{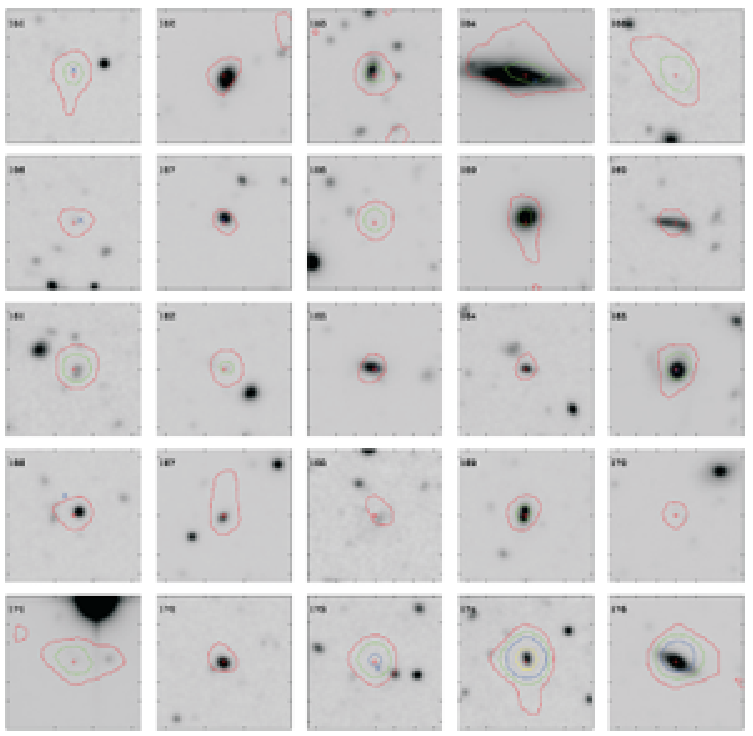,width=\textwidth}
\resizebox{0.8\hsize}{!}{\includegraphics{f11.eps}}
\end{center}
\caption{\label{WFI_IDs} See Figure 5 caption. } 
\end{figure}

\begin{figure}[p] 
\begin{center} 
\includegraphics{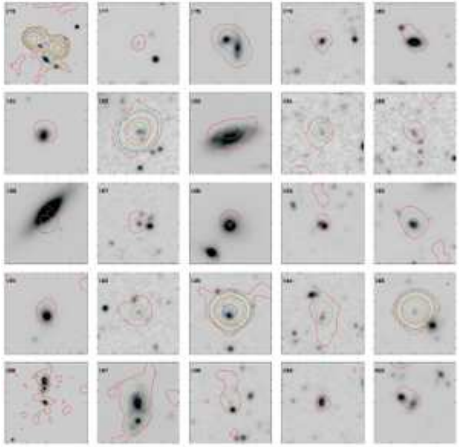,width=\textwidth}
\resizebox{0.8\hsize}{!}{\includegraphics{f12.eps}}
\end{center}
\caption{\label{WFI_IDs} See Figure 5 caption.  RID~176 is 50 arcsec on a side and RID~196 is 50 arcsec. } 
\end{figure}

\begin{figure}[p] 
\begin{center} 
\includegraphics{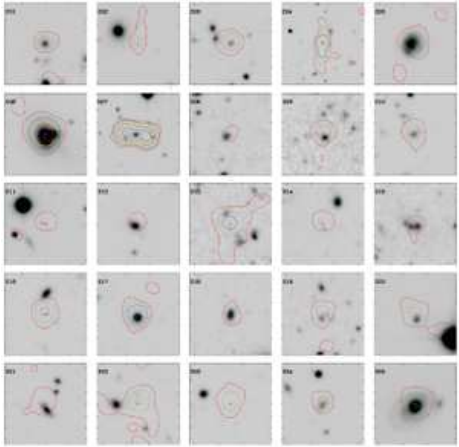,width=\textwidth}
\resizebox{0.8\hsize}{!}{\includegraphics{f13.eps}}
\end{center}
\caption{\label{WFI_IDs} See Figure 5 caption.  RID~202 is 25 arcsec on a side, RID~204 is 40 arcsec, and RID~207 is 40 arcsec.} 
\end{figure}


\begin{figure}[p] 
\begin{center} 
\includegraphics{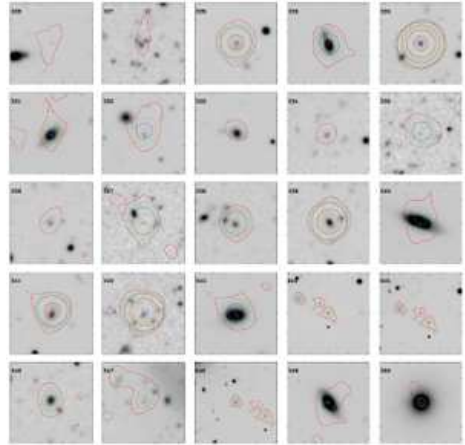,width=\textwidth} 
\resizebox{0.8\hsize}{!}{\includegraphics{f14.eps}} 
\end{center}
\caption{\label{WFI_IDs} See Figure 5 caption. RID~244, 245, and 248 are 50 arcsec on a side.
}
\end{figure}

\begin{figure}[p] 
\begin{center} 
\includegraphics{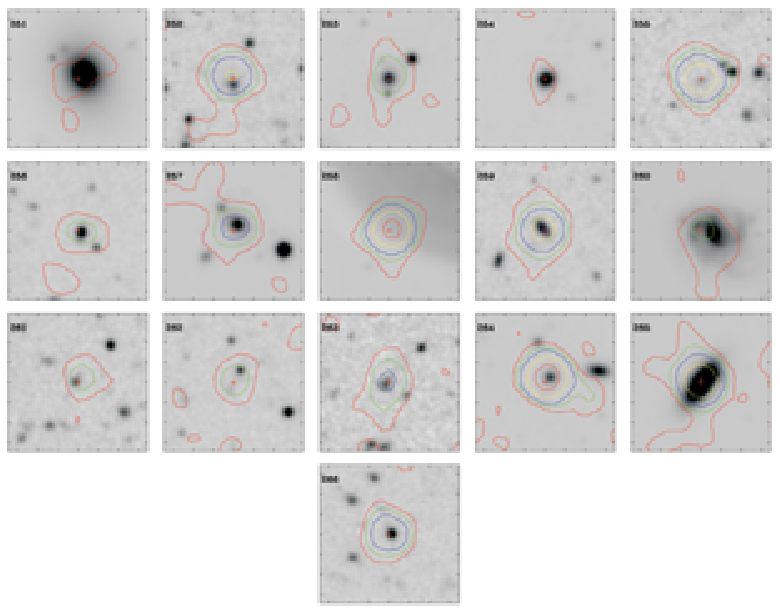,width=\textwidth}
\resizebox{0.8\hsize}{!}{\includegraphics{f15.eps}}
\end{center}
\caption{\label{WFI_IDs} See Figure 5 caption. } 
\end{figure}



\subsection{\it The IR Data \label{IR}}

GOODS/ISAAC: As part of the GOODS survey an area of 131 arcmin$^{2}$ has been
imaged with ISAAC at the VLT in three bands: J (25.2, H(24.7), Ks(24.7, AB 5$\sigma$ for 1 arcsec
area) (J. Retzlaff, in preparation).  \footnote{See
\url{http://archive.eso.org/archive/adp/GOODS/ISAAC\_imaging\_v2.0/index.html}}

EIS/SOFI: A larger area (380 arcmin$^{2}$) has been imaged with SOFI/NTT as part
of the ESO Imaging Survey. Two filters have been used: J (23.4 and Ks (21.4), AB 5$\sigma$ for
2.0 arcsec apertures). \citep{ols2006}\footnote{See also
\url{http://www.eso.org/science/goods/imaging/products.html}}

GOODS/IRAC: An area of 160 arcmin$^{2}$ in the CDF-S has been imaged by the IRAC
camera on board of the Spitzer telescope as part of the GOODS survey. Data have
been taken in the four channels: 3.6, 4.5, 5.8 and 8.0 micron for an exposure of
24 hours.  \footnote{IRAC images can be found at:
\url{http://data.spitzer.caltech.edu/popular/goods/20041027\_enhanced\_v1}}

GOODS/MIPS:  A 30' x 30' area coincident with the E-CDF-S was observed by Spitzer
  as part of the Far-Infrared Deep Extragalactic Legacy (FIDEL) program.
  The data currently available are based on a typical exposure times of 2400 to
  of 4800s at 24 and 70 microns respectively.  \footnote{See
\url{http://ssc.spitzer.caltech.edu/legacy/fidelhistory.html}}

Optical or IR counterparts were found for 254 of the 266 cataloged radio sources.  Three radio sources have no optical or IR counterpart, while for the other nine sources, there is more than one object near the radio position and the identification is ambiguous.  As noted by 
\cite{ric99}, a small fraction of sub millijansky radio sources have
no apparent optical counterpart, even in the deepest optical exposures.  RID 97 which is a relatively strong 1.5 mJy radio source, is an extreme example with no apparent optical, IR, or X-ray counterpart.  RID 97 and RID 92 which lies 38'' away and is 3.7 mJy could be part of double radio galaxy.  However, RID 92 is coincident with an X-ray source and appears to be firmly identified with a 22.5 mag galaxy so this interpretation is unlikely.  We cannot exclude the possibility, that RID 97 is part of RID 92, although there is no evidence of a connection between the two sources.

Spectroscopic Data: Redshifts are available for 186 of the 266 radio
sources;  108 are measured spectroscopically \citep{SBH04} and 78 are photometric taken from the COMBO-17 
survey \citep{2004A&A...421..913W}. Also, see Paper II.  For sources with R$<24$ 
the COMBO-17 survey provides photometry in 17 passbands from 350 to 930 nm which gives photometric redshifts typically accurate to $\sim 2\%$ for the brighter galaxies ranging to $\sim 10\%$ accuracy for the fainter ($R \geq 24$) galaxies.

\section{RADIO PROPERTIES\label{properties}}


\subsection{\it Extended Radio Sources \label{extended}} 

Most of the radio sources
detected by the VLA are compact and are either unresolved or barely resolved by
our 3.5 arcsecond beam. They are likely either AGN or small regions of
enhanced star formation.  With the excellent positional accuracy of both the
Chandra and the VLA observations, the optical or near IR counterparts of these sources are
generally uniquely defined by positional coincidence.  About ten percent of the sources are well resolved by the VLA with angular sizes greater than 5 arcseconds.  Most of these resolved sources have weak low surface brightness extensions, but ten (RID 7, 18, 23, 88, 113, 140, 148, 176, 207, and 248) have multiple components with a centrally located optical counterpart and radio luminosity (where known) characteristic of the more powerful classical radio galaxies.


\vspace{2.5cm}

\subsection{\it Comments on Individual Sources}

In this section we comment on individual source structure and their counterparts observed in other wavelength bands.  Further detail on the optical counterparts is given in Paper II.  

RID 1 \& 4~~may be associated.

RID 5 \& 6~~may be associated.

RID 7~~is a double source with an optical counterpart and X-ray source coincident with the
compact central component.

RID 8~~The optical counterpart is part of a small group.

RID 16~~There is evidence of radio emission from a faint galaxy located about 3.5 arcsec to the north west.

RID 18~~May be a blend of two independent sources, only one of which (A) has an optical counterpart.

RID 19~~is located about 15 arcsec from a bright star.

RID 23~~This bright bent double lobe radio source has no visible optical
counterpart.

RID 25~~The optical counterpart is visible on a z band GEMS image.

RID 34~~This compact low luminosity radio source is located close to the center of a nearly edge-on spiral seen in the z band GEMS image.

RID 44~~This flat spectrum radio source is located 1.2 arcsec from a very red galaxy.

RID 50~~The peak of the radio emission falls 1.1 arcsec from a very red R=24 mag galaxy, but an extension of the radio source is very close to a bright spiral galaxy and may be an independent radio source.

RID 53~~The X-ray source XID 121 is 4.4 $\pm$ 1.9 arcsec away from the radio
source and coincident with another galaxy with weak radio emission. 

RID 58~~A faint galaxy is seen on the GEMS ACS z-band image at the radio position.  See Paper II.

RID 60~~The faint extension to the north east may be related to the bright spiral galaxy.
 
RID 63~~The radio source may be associated with a compact group of galaxies. 

RID 70~~No visible R band counterpart, but detected in Spitzer IRAC bands.

RID 71~~The optical counterpart appears to be a member of a close binary pair.

RID 73~~is a fairly bright double lobe radio source with a very red optical/IR counterpart.

RID 78~~The radio source may be associated with a compact galaxy group.

RID 88~~The optical and X-ray counterparts are coincident with the central component of this double lobe radio source.

RID 97~~has no apparent X-ray or optical counterpart in any band out to 8 microns.  RID 97 has an extreme value of the ratio of radio to optical or X-ray luminosity.

RID 113~~is a luminous extended double lobe source.  The central component is
identified with a mag 19.4 galaxy.  The X-ray counterpart (XID 249) is located 3.9
arcsec away near the outer part of the galaxy.  A second X-ray source (XID 527)is
coincident with the south west lobe of the radio source.

RID 126~~The counterpart is detected in ISAAC J band.

RID 140~~The optical counterpart of this double lobe radio galaxy is coincident with the weak central component A.  RID 141 and 142 are remarkably close in projection but have a different redshift than RID 140 (0.22)

RID 141~~and 142 have the same redshift (0.08) and appear to be part of the same cluster.

RID 144~~is seen in the H SOFI band.

RID 148~~is a powerful double lobe radio galaxy.

RID 151~~is seen at 3.6 microns.

RID 154~~The weak radio emission covers the full extent of the optical image of this edge on spiral galaxy.

RID 167~~The brightest component, 167A of this extended source is coincident with a mag 21.94 galaxy.

RID 169~~The radio source is located between a pair of galaxies separated by 1.3 arcsec.

RID 171~~The optical counterpart is too close to a bright star to derive a meaningful magnitude.

RID 173~~The optical counterpart is easily visible on the ACS image.

RID 176~~A red galaxy located near the centroid of this extended double radio source is visible in the 3.6 micron Spitzer image.  The X-ray source is coincident with the galaxy, but there is no associated radio component at this position.

RID 178~~Components A and B are each coincident with a different galaxy at somewhat different redshifts.  They are probably independent sources.  The X--ray feature is probably associated with RID 178B.  

RID 180~~The X-ray source XID 646 is 5.2 $\pm$ 2.1 arcsec from the centroid of
the radio source, but lies within the boundaries of both the radio source and the
apparent optical counterpart.

RID 183~~This extended radio source covers the optical image.

RID 186~~Although the radio source is separated by 3.1 $\pm$ 1.7 arcsec from
the center of the optical image, it lies well within the galaxy and is likely from a star forming region in the galaxy.  The X-ray source, however, is located near the center of the galaxy and is probably unrelated to the radio source.

RID~~196 \& 197 are coincident with a pair of galaxies at z=0.24.  An extension of RID 197 to the south is coincident with another pair of galaxies at unknown redshift, but which are probably part of the same group.

RID 207~~The central component of this classical double lobe radio source is identified with a R = 20.4 E galaxy.

RID 213~~No optical counterpart for this 275 $\mu$Jy source visible down to R mag 25.9 in the ACS GEMS image.

RID 216~~A red galaxy is seen on the spitzer 3.6 $\nu$ image but no optical counterpart is visible down to R = 25.9.

RID 215~~The radio source is located between two galaxies with uncertain redshifts.

RID 227~~The radio source is coincident with a small group of galaxies. 

RID 244, 245, and 248~~is probably a single source with an optical counterpart associated with the NE component.  

RID 247~~may be two independent sources with separate optical counterparts.

RID 258~~The counterpart is visible in a K band SOFI image.

\subsection{Comparison with the ATCA Survey}

The CDF-S has also been observed at 1.4~GHz using the Australia Telescope Compact
Array (ATCA).  The ``Australia Telescope Large Area Survey,'' or ATLAS\citep{nor2006} 
surveyed a 3.7 deg$^2$ area including the CDF-S using a mosaic of 28 ATCA pointings. The 
seven pointings that were centered on the CDF-S (one central pointing surrounded 
by a hexagonal ring of six pointings) were more tightly spaced and received longer 
integrations, and were used to create the radio source list contained within the GOODS 
coverage area \citep{A06}. \citet{nor2006} provides a master catalog for the 
full ATLAS observations, including sources within the GOODS area. 

\begin{figure}[p]
\epsscale{0.9}
\plotone{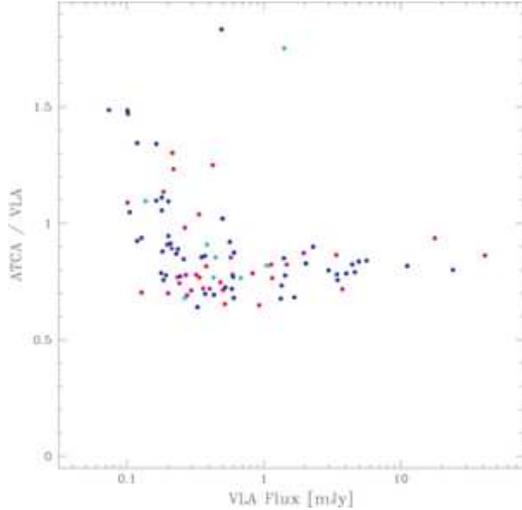}
\caption{\label{fig-norris}Comparison of flux densities from Table 1 with corresponding sources in the 
\citet{nor2006} ATLAS catalog. The y-axis shows the ratio of ATCA/VLA flux density values 
from the \citet{nor2006} catalog and the x-axis the VLA values.  
For sources stronger than 200 $\mu$Jy the VLA values appear about 20 percent lower than 
the ATCA values while at lower flux densities there is a lot of scatter.  The red points represent 
sources which are resolved with both the ATCA and the VLA; purple points, resolved only by the VLA; 
green points, resolved only by the ATCA, and blue points, unresolved by both the VLA and the ATCA.
The large outlier near 1.5 mJy is RID 264. The flux density given by \citet{M08} agrees with that in Table I.  Comparison of the other outlier, RID 258, with a VLA flux density of 0.47 mJy with the \citet{M08} catalog suggests that RID 258 may be variable.
}
\end{figure}

We have compared the catalog of Table~\ref{catalog} with the ATLAS catalog which lists individual
components of extended sources, so that a double lobed source is cataloged
as two separate entries. A direct matching of the two data sets produces 126 ATLAS sources that
coincide with entries from Table~\ref{catalog}, where the Table~\ref{catalog} counterparts may be individual
components or complete sources. A straight comparison of the flux densities from the
two surveys suggests that the flux density scale of the ATLAS observations is $16\pm3\%$ 
lower than that of the VLA observations. In this comparison, we have used the 
integrated flux density values for the ATLAS data as these
proved to be better matched to the VLA data than either the peak flux densities or
a simple scheme of adopting the peak flux densities for unresolved sources and the 
integrated flux densities for resolved sources. Of course, the poorer resolution of the 
ATLAS data (11$^{\prime\prime} \times 5^{\prime\prime}$) makes such a simple direct matching 
flawed; separate VLA sources may be blended into a single ATLAS source, the VLA 
observations may resolve out flux density detected in the ATLAS observations, or actual 
mismatches could occur. Additionally, the measured flux densities of fainter sources
can depend on the details of the fitting algorithm used to derive their values. To account for these 
factors, we performed iterative sigma clipping on the matched list to remove outliers. 
The clipping converged when 94 sources had matches between the VLA and ATLAS catalogs, 
with the ATLAS catalog having flux densities $20\pm1\%$ below those of the Table 1 values. 
Results for the matched catalogs are shown in Figure \ref{fig-norris}, where it can be 
seen that the two data sets can be reconciled by a simple 20\% calibration error. 
This possible discrepancy in the flux 
density scale of the ATLAS catalog was noted in \citet{nor2006}, where comparison of
the ATLAS values with the separate reductions of the CDF-S data \citep[Koekemoer et al., in 
preparation]{A06} and with the NVSS \citep{1998AJ....115.1693C} indicated a 10\% - 20\% discrepancy.  However, a re imaging of the ATCA data by E. Middleberg, (private communication) has brought the ATCA flux densities to within about 5\% of the VLA values.


\section{Radio -- X-ray Relations \label{supplementary}}


Table~\ref{catalog} includes only radio sources with peak amplitudes at 1.4 GHz greater
than 5 times the rms nose, or $43~\mu$Jy in the image before correction for the radial attenuation shown in Fig.~\ref{sensitivity}.  Of the 266 cataloged radio sources listed in Table~\ref{catalog}, 52 have X-ray counterparts in the CDF-S 1Ms exposure (in the solid angle covered with more than 25 percent of the maximum exposure time) and 37 are found in the complementary area covered by the E-CDF-S, for a total of 89 X--ray/Radio matches.

There are about $10^5$ independent beam areas covered by our radio image, but our 5-sigma cutoff ensures that there is
less than a 0.5 percent probability that any of the 266 radio source listed in Table~\ref{catalog} is simply due to a
random noise fluctuation. However, at the position of the cataloged 
X-ray sources (or any other set of selected positions) we can make meaningful measurements of the radio flux density down
to much lower values corresponding to 2 or 3 times the rms noise.  The 1.4 and 5 GHz radio emission from each of the 338 cataloged CDF-S X-ray sources (whether a formal 5-sigma radio detection or not) was
determined from the flux density measured in the clean radio image at the location of each Chandra
X-ray source. These measured radio flux densities are
tabulated in Table~\ref{radio_xray} along with their measured rms error.  The values given in Table~\ref{radio_xray} refer only to the peak value in our 3.5 arcsecond beam.  Thus, if the radio source is extended, or if the X-ray position is in error, the values listed in Table~\ref{radio_xray} will be less than that given in Table~\ref{catalog}.  Thirty additional X--ray sources were found with a SNR between 3 and 5 in the radio image.

\begin{figure}[ht]
\includegraphics{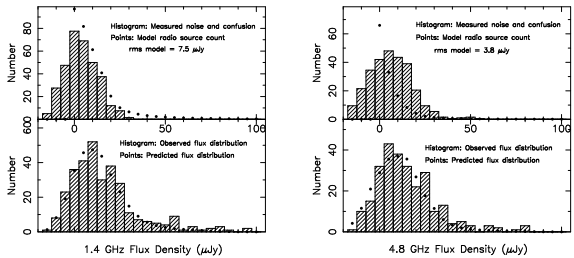}
\vskip 3 cm 
\caption{\label{radioXray} Radio emission from X-ray sources. The left-hand panels
are associated with the 1.4 GHz radio emission; the right-panels with
the 4.8 GHz radio emission.  The top panels shows the histogram of the
measured noise and confusion at each frequency.  The plotted points
are the assumed radio count for the x-ray sample.  The bottom panelsshows the histogram of the observed radio flux distribution for the338 sources in the x-ray sample.  The plotted points are thepredicted radio flux distribution obtained by convolving the assumedsky model and noise distribution shown in the top panels.
}  
\end{figure}

If the radio data were limited only by thermal noise, the average noise level could be reduced by averaging measurements made at a number of selected X-ray positions.  However, in case of long VLA integrations at 20 cm, confusion from unresolved weak sources is significant.   Therefore, in order to examine the distribution of 
the faint radio emission from the Chandra X-ray
sources, we have compared the observed distribution of measured radio flux
densities in Table~\ref{radio_xray} with that expected from a model
count of faint sources including the contributions of noise and confusion.  The results, which are
illustrated in Fig.~\ref{radioXray} show that the cataloged Chandra X-ray sources do have detected radio
emission that is systematically larger than that expected from noise and confusion from
unrelated background sources.

The plots on the left and right side of Fig.~\ref{radioXray} are associated with
the 1.4 GHz and 4.8 GHz data, respectively.  The histograms in the top
panels show the distribution of only noise plus confusion determined by
measuring the radio flux density at four locations, each separated by
$15"$ in the cardinal directions from the cataloged X-ray sources. The average of these measurements over the 338 ''blank'' fields should
accurately reflect the distribution of noise and confusion.

To model the faint radio source distribution associated with the X-ray
sources, we assumed a power law source count with a
differential slope of $-2.1$ as obtained for the stronger radio sources.
The only free parameter of
this distribution is a scale factor which is determined by the
number of radio sources in the sample.  The best fit count of this form is shown by
the plotted points in the upper panels.  It rises steeply toward low flux
densities and is zero at negative flux densities (the number of sources shown in the $-5~\mu$Jy to $+5~\mu$Jy bin
reflects the number of sources only between 0 to $+5~\mu$Jy).

The predicted observed flux density distribution obtained from convolving the above count with the measured noise and
confusion is shown by the plotted points in the lower panels along with the
histogram of the observed flux density from Table~\ref{radio_xray}.  The mean radio flux density for the X-ray sources is
$7.5\pm 0.7~\mu$Jy at 1.4 GHz and $3.8\pm~0.9~\mu$Jy at 4.8 GHz.
These values have been calculated only for the radio flux
densities that are less than $43~\mu$Jy; that is, only for those sources
below the 5-sigma detection level used to compile Table 1.

The average spectral index of the 1.4 GHz and 4.8 GHz faint radio emission
from the X-ray sample is $0.55\pm 0.20$, in
agreement with that of the detected sources.  

\section{Source Count}

The complete radio sample contains 198 sources with a peak signal to noise $>5.0$, within $15'$ of the field center.  The distribution of the flux densities of these sources is given in Table~\ref{count}.
Columns 1, 2 and 3 give the flux-density bin parameters, and column 4 gives the number of observed sources with an integrated flux density within each bin.  We then used a Monte-Carlo simulation to determine the number of sources from a Euclidean count of the form 
$dn(S)=S^{-2.5}dS$, where $dn$ is the number of sources (Jy)$^{-1}$ (sr)$^{-1}$ between flux density $S$ and $S+dS$ that would be detected above a peak image flux density of 43~$\mu$Jy.  The simulation steps included: 1). The $15'$ region was randomly populated with sources having the above Euclidean number density; 2) The primary beam attenuation, beam smearing and integration time smearing (see Fig.~\ref{sensitivity}) were applied to each source; 3) Each source was given an angular size according to the distribution shown by \citet{fom06} (see Fig. 9); and 4) Gaussian noise of $8.5~\mu$Jy rms was added to the peak intensity of each source.

In columns 5 and 6 we give the results of this simulation.  Column 5 shows the fraction of sources, F, in the sky within the 15' radius, that would be detected by the observations.  Column 6 gives the number of sources expected from the Euclidean count after instrumental, angular-size and noise effects were considered.  Column 7 lists the normalized count, the ratio of column 4 to column 6.  The estimated error in column 7 is equal to the count value divided by the square-root of the number of sources in column 4 and does not include uncertainties in the source count modeling, particularly the assumed angular distribution of the sources and the effects of non-random errors in the final image.  It is possible that the lower flux density bin in error by perhaps 30\% of its value.

The best fit to a power law of the count in the CDF-S between 43 and 2000 $\mu$Jy is

$$  N(>S) = (0.091\pm 0.014) (S/200)^{-1.10\pm 0.13}  $$

\noindent where N($>$S) is the number of sources (arcmin)$^{-2}$ with a flux
density greater than $S~\mu$Jy. The CDF-S source count can also be described in differential form with a best fitting power law of the form 

 $$dn(S) = 106~S^{-2.1}dS,$$

\noindent where dn is the
number of sources between flux density S and dS (in Jy) per Jy
interval per ster.  At a flux density of $50~\mu$Jy, this corresponds
to a count of 

$$dn(S)=0.0092~[S/50]^{-2.1} $$

\noindent sources per $\mu$Jy per(arcmin)$^{-2}$ and where S is in $\mu$Jy.
 
The CDF-S count agrees with the well known 
flattening (closer to the Euclidean slope) of the normalized differential count
below about $300~\mu$Jy, whereas above $500~\mu$Jy the steep
convergence of the count is apparent in all surveys.  The observed 
number density above $50~\mu$Jy in the CDF-S is 0.41 sources (arcmin)$^{-2}$, corresponding to a
mean separation between sources of about $90"$.  If the CDFS count is extrapolated to
$1~\mu$Jy, the average separation of sources would be about $6"$, but
based on the other deep surveys, the true number may be somewhat larger and the 
mean separation correspondingly smaller.

\begin{figure}[ht]
\begin{center}
\includegraphics{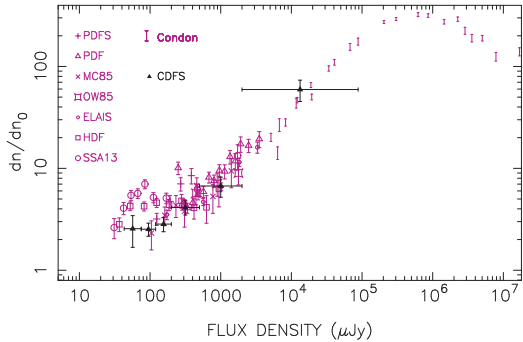,width=4in}
\rotate
\resizebox{0.9\hsize}{!}
{\includegraphics{f18.eps}}
\end{center}
\caption{\label{allcount}The 1.4 GHz (20 cm) source count from the complete sample
of the CDF-S field is shown as black triangles along with the counts 
from other deep surveys.  References are as given in the text. The points for the highest 
flux density values shown without symbols are based on a compilation from \cite{C84}.
}
\end{figure}

The CDFS count is plotted in Fig.~\ref{allcount}, along with the results
from other deep surveys of comparable sensitivity.  References to
other counts are: PDFS and PDF \citep{1998MNRAS.296..839H}, MC85
\citep{mit85}, OW84 \citep{oor85}, ELIAS \citep{cil00}, HDF
\citep{ric00}), and SSA13 \citep{fom06}. The highest flux density
points are based on a compilation from \cite{C84}.

As is well known, there are three regions of the radio source count.  The
rise in the counts, with respect to a Euclidean count, from the
brightest sources down to $\sim$1~Jy is caused by the strong evolution of the luminous quasars and radio galaxies that dominate
this source population.  The drop-off in the count from near 1 Jy to a few millijanskys,
where quasars and radio galaxies still dominate, is caused
predominantly by the redshift-cutoff above z=3 in the density of these
luminous radio galaxies.  

The flattening of the normalized count (although less pronounced with
the CDF-S field), which is observed over an order of magnitude in flux density
(35 to 300 microJy) in many deep survey observations, is produced
by a separate population of objects which is dominant in the
submillijansky count because there are very few luminous quasars and
galaxies found above a redshift of about 5.  As discussed by
P. Padovani et al. (in preparation) the sub-millijansky population
appears to be due synchrotron radiation from a mixture of star forming
regions and low luminosity AGN.  Further study of the submicrojansky radio population
with the new X-ray, IR, and optical material should lead to a better understanding of the relation between star
formation and massive black holes and their evolution.

Finally, we note that the field-to-field scatter appears
to increase below about 400 $\mu$Jy where the observed number density of source in the
CDF-S is somewhat lower than the other surveys.  For example,
the SSA13 field \citep{fom06} contains 510 sources above
$43~\mu$Jy in a $15'$ radius, compared with 198 above this level from
the CDF-S field. Hence, the average count in the CDF-S is more than a factor of 2
less than that for the SSA13 field.  Both surveys were made with the
VLA using similar analysis procedures so the difference cannot simply be explained
by instrumental errors.  This supports suggestions that the scatter in the
density of sources near the $\sim 100~\mu$Jy level among the deep
field regions may due to
real cosmic variance on scales of a few square degrees. However, more careful comparison of the different survey fields using consistent procedures for instrumental corrections is needed to verify the extent that true csomic variance is important on degree scales.

\section{SUMMARY \label{summary}}

The VLA survey of the CDF-S and E-CDF-S has cataloged 266 radio sources above a limiting flux density of $43~\mu$Jy in the most sensitive part of the field.  Typical radio position accuracy of better than 1 arcsecond and multiwavelength imaging allowed optical or NIR counterparts to be identified for more than 95 percent of the cataloged source with spectroscopic or photometric redshifts available for about 70 percent of the radio sources.  Although for most of the unidentified sources the lack of a unique OIR counterpart is due to ambiguities resulting from multiple faint galaxies within the radio position uncertainty, there are a few empty fields with no apparent counterpart down to limiting magnitudes as faint as mag 25.9. 

Most of the radio sources are unresolved or only barely resolved with the 3.5 arcsecond VLA beam at 20 cm, although about ten percent are well resolved with angular sizes greater than 5 arcseconds.  Most of the resolved sources have weak low surface brightness extensions.  Ten sources, all of which have a total flux density greater than 1 mJy, have multiple components characteristic of classical radio galaxies, while the submillijansky population is thought to be due to synchrotron radiation from regions of active star formation and low luminosity AGN.

Eighty-nine radio sources in our complete radio catalog were found to have X-ray counterparts in either the 1
Megasecond Chandra catalog or in the E-CDF-S.
In addition to the cataloged radio sources, we also give the measured radio emission from X-ray sources found in the CDF-S catalog.  An additional 30 CDF-S X-ray sources were detected at 20 cm were found with SNR between 3-sigma and 5-sigma. 

Below a few hundred microjanskys, the differential radio source count for the CDF-S approaches the ''Euclidian" slope of -2.5 corresponding to the evolving population of starforming regions and low luminosity AGN.  The scatter among different surveys is large, and may be due, at least in part, to cosmic variance.  Although, the uncertainties in the instrumental corrections for the weaker sources, especially for resolution, are also large and may account for some of the discrepencies among observers, it is hard to explain the observed large scatter at the few hundred microjansky level as the result of instrumental corrections.  However, we note the significant difference in the flux density scales of the VLA and ATCA observations of the same CDF-S. 

Padovani et al. (in preparation) use these data to discuss the radio/X-ray/OIR relationships and the evidence for contribution to the microjansky radio emission from AGN as well as from star formation  

\cite{M08} has used the VLA at 1.4 GHz to cover the full E-CDF-S with a sensitivity $\sim$~$8~\mu$Jy rms, which will be complemented by deeper new Chandra X-ray and Spitzer IR observations as well as new VLA 6 cm observations.  Later, ALMA with its unprecedented sensitivity and resolution at sub-millimeter wavelengths will help to unravel the nature of the sub-millijansky radio source population and the relation between AGN and star forming activity.

\acknowledgements

We thank the referee for constructive suggestions which ahve helped to improve the clarity of the presentation.  The VLA is a facility of the National Radio Astronomy Observatory which is
operated by Associated Universities, Inc., under a cooperative agreement with the
National Science Foundation.  Part of this work was done while VM was at the Max Planck 
Institut f\"ur extraterrestrische Physik, Giessenbachstrasse 
1, D--85748 Garching, Germany.
PT acknowledges the financial contribution from contract ASI-INAF I/023/05/0.  We acknowledge the ESO/GOODS project for the ISAAC and FORS2 data obtained using the Very Large Telescope at the ESO Paranal Observatory under Program ID(s): LP168.A-0485, 170.A-0788, 074.A-0709, and 275.A-5060.

\clearpage




\begin{deluxetable}{lrrrrcccrrrrccc}
\tablecolumns{15}
\rotate
\tablewidth{8.8in}
\tabletypesize{\scriptsize}
\tabcolsep3.8pt
\tablecaption{\label{catalog}CDF-S Radio Source Catalog}
\tablehead {
  \colhead{(1)} &
  \colhead{(2)} &
  \multicolumn{2}{c}{(3)} &
  \colhead{(4)} &
  \colhead{(5)} &
  \colhead{(6)} &
  \colhead{(7)} &
  \colhead{(8)} &
  \colhead{(9)} &
  \colhead{(10)} &
  \colhead{(11)} &
  \colhead{(12)} &
  \colhead{(13)} &
  \colhead{(14)}  \\

  \colhead{Source} &
  \colhead{No} &
  \colhead{RA} &
  \colhead{DEC} &
  \colhead{SNR} &
  \colhead{S$_{1.4}$} &
  \colhead{$\theta$} &
  \colhead{S$_{4.8}$} &
  \colhead{$\alpha$} &
  \colhead{z} &
  \colhead{R} &
  \colhead{X-ID} &
  \colhead{S$_x$} &
  \colhead{L$_x$} &
  \colhead{L$_r$} \\

   & & & & &
  \colhead {($\mu$Jy)} &
  \colhead {(arcsec)} &
  \colhead {($\mu$Jy)} \\
}
\startdata
J033111.7$-$273144 N &   1 & $03~31~11.70\pm 0.02$ & $-27~31~44.3\pm 0.4$ &    49.6 & $   3420\pm    93$ & $ <1.5$ & $              $ & $              $ &       & 16.79 &       &        &$      $&        \\  
J033113.9$-$273910 N &   2 & $03~31~13.96\pm 0.03$ & $-27~39~10.3\pm 0.3$ &    48.1 & $   1430\pm    28$ & $ <1.5$ & $              $ & $              $ &       & 23.13 &       &        &$      $&        \\  
J033114.2$-$274844 N &   3 & $03~31~14.20\pm 0.03$ & $-27~48~44.3\pm 0.3$ &     4.9 & $    149\pm    25$ & $ 3.2 $ & $              $ & $              $ &       & 25.12 &       &        &$      $&        \\  
J033114.4$-$273907 N &   4 & $03~31~14.42\pm 0.06$ & $-27~39~07.1\pm 0.6$ &     6.1 & $    142\pm    25$ & $ <2.5$ & $              $ & $              $ &       & 18.04 &       &        &$      $&        \\  
J033114.5$-$275546 N &   5 & $03~31~14.50\pm 0.03$ & $-27~55~46.3\pm 0.3$ &    11.4 & $    360\pm    28$ & $ 2.7 $ & $              $ & $              $ &       & 23.40 &       &        &$      $&        \\  
J033114.8$-$275543 N &   6 & $03~31~14.87\pm 0.06$ & $-27~55~43.4\pm 0.6$ &     5.1 & $    110\pm    23$ & $ <2  $ & $              $ & $              $ &       & 23.28 &       &        &$      $&        \\  
J033115.0$-$275518 N &   7 & $03~31~15.06\pm 0.05$ & $-27~55~18.4\pm 0.6$ &   142   & $  11920\pm   237$ & $ 20  $ & $              $ & $              $ &  0.49 &       &     &        &$ 43.66$&  24.99 \\  
                     &   7A &$03~31~14.00\pm 0.02$ & $-27~55~19.3\pm 0.1$ & $  108   $ & $ 7190 \pm  156$ & $7.0 $ &                 &                  &      &       &      &        &        &         \\
\enddata
\end{deluxetable}


\begin{deluxetable}{rrrr|rrrr|rrrr}
\tablecolumns{12}
\tablewidth{6.5in}
\tabletypesize{\footnotesize}
\tabcolsep4.0pt
\tablecaption{\label{radio_xray}Radio Flux Densities for X-ray Sources} \tablehead {

  \colhead{X-NO} &
  \colhead{S$_x$} &
  \colhead{S$_{1.4}$} &
  \colhead{S$_{4.8}$} &
  \colhead{X-NO} &
  \colhead{S$_x$} &
  \colhead{S$_{1.4}$} &
  \colhead{S$_{4.8}$} &
  \colhead{X-NO} &
  \colhead{S$_x$} &
  \colhead{S$_{1.4}$} &
  \colhead{S$_{4.8}$} \\
}
\startdata
001 &     6.0 &    46$\pm$ 10  &    40$\pm$ 21  & 002 &     4.9 &    18$\pm$ 11  &    22$\pm$ 20  & 003 &     3.6 &     0$\pm$ 10  &    23$\pm$ 17  \\
004 &     9.4 &    12$\pm$ 10  &   -18$\pm$ 13  & 006 &    10.2 &    54$\pm$ ~9  &    30$\pm$ 15  & 007 &    23.5 &    34$\pm$ 12  &                \\
008 &     9.6 &    22$\pm$ 10  &   -22$\pm$ 14  & 009 &    16.2 &    22$\pm$ 10  &   -24$\pm$ 14  & 010 &     9.8 &    10$\pm$ ~9  &     9$\pm$ 11  \\
011 &    17.7 &    22$\pm$ ~9  &    22$\pm$ 13  & 012 &     6.7 &    -7$\pm$ ~9  &    -7$\pm$ 11  & 013 &    14.3 &    24$\pm$ 10  &    13$\pm$ 10  \\
015 &     8.4 &     6$\pm$ ~9  &    16$\pm$ ~8  & 017 &     2.7 &    -7$\pm$ ~9  &    15$\pm$ ~9  & 018 &    33.6 &   158$\pm$ ~9  &    75$\pm$ ~8  \\
019 &    19.4 &     0$\pm$ ~9  &    -6$\pm$ ~9  & 020 &     4.2 &    19$\pm$ ~8  &    12$\pm$ ~9  & 021 &     1.4 &    12$\pm$ ~9  &    25$\pm$ ~7  \\
022 &     9.5 &     7$\pm$ ~8  &    12$\pm$ ~9  & 023 &     4.0 &     8$\pm$ ~9  &    -7$\pm$ ~8  & 024 &     5.6 &   -10$\pm$ ~8  &     4$\pm$ ~7  \\
025 &    12.5 &   139$\pm$ ~9  &    52$\pm$ 11  & 026 &     3.4 &    28$\pm$ ~8  &    -4$\pm$ ~8  & 027 &     7.6 &    49$\pm$ ~8  &     6$\pm$ 10  \\
\enddata
\end{deluxetable}

\begin{deluxetable}{rrrcrcc}

\tablecolumns{7}
\tabletypesize{\normalsize}

\tablecaption{\label{count}CDF-S Differential Radio Count}

\tablehead {

\colhead{(1)} &

   \colhead{(2)} &
   \colhead{(3)} &
   \colhead{(4)} &
   \colhead{(5)} &
   \colhead{(6)} &
   \colhead{(7)}  \\
   \colhead {S$_{min}$} & 
   \colhead {S$_{max}$} &
   \colhead {S$_{avg}$} &
   \colhead {N}     &
   \colhead {F}      &
   \colhead {N$_{E}$}   &
   \colhead {dn/dn$_0$}  \\
   \colhead{$\mu$Jy} &
   \colhead{$\mu$Jy} &
   \colhead{$\mu$Jy} \\
  }

\startdata

     43 &   75 &   62.2 &  34 &  0.19 & 13.2 & 2.58$\pm$ 0.45 \\
     75 &  120 &   95.2 &  46 &  0.61 & 18.2 & 2.52$\pm$ 0.38 \\
    12 &  200 &  148 .8 &  40 &  0.91 & 13.9 & 2.88$\pm$ 0.45 \\
    200 &  500 &  286.3 &  40 &  0.98 &  9.4 & 4.23$\pm$ 0.68 \\
    500 & 2000   &  851.6 &  20 &  1.00 & 3.04 & 6.58$\pm$ 1.51 \\
    $>$2000 &        & 5200   &  18 &  1.00 & 0.33 & 55  $\pm$ 13   \\

\enddata

\tablecomments{Col. 1: the flux density bin lower limit; Col. 2: the flux density bin upper limit; Col. 3: the weighted average flux density in bin; Col. 4: the number of observed sources in bin; Col. 5: the fraction of observed area in which the weakest sources in this bin could be detected; Col. 6: the number of observed sources expected if the count is Euclidean after correcting for instrumental errors; Col. 7: normalized differential source count}

\end{deluxetable}

\end{document}